\documentclass[twocolumn,prl,superscriptaddress]{revtex4}
\usepackage{graphicx} 
\usepackage{dcolumn}  
\usepackage{bm}       
\usepackage{amsfonts}
\usepackage{amssymb}

\begin{document}

\title{Confinement of hydrodynamic modes on a free surface and their quantum
analogs}

\author{M.  Torres\footnote{Electronic address: manolo@iec.csic.es}}
\affiliation{Instituto de F\'{\i}sica Aplicada, Consejo Superior de
  Investigaciones Cient\'{\i}ficas, Serrano 144, 28006 Madrid, Spain.}

\author{J. P. Adrados}
\affiliation{Instituto de F\'{\i}sica Aplicada, Consejo Superior de
  Investigaciones Cient\'{\i}ficas,
  Serrano 144, 28006 Madrid, Spain.}

\author{P. Cobo}
\author{A. Fern\'andez}
\affiliation{Instituto de Ac\'ustica, Consejo Superior de
  Investigaciones Cient\'{\i}ficas, Serrano 144, 28006
  Madrid, Spain}

\author{C. Guerrero}
\affiliation{Laboratorio de F\'{\i}sica de Sistemas Peque\~nos y
  Nanotecnolog\'{\i}a, Consejo Superior de Investigaciones
  Cient\'{\i}ficas, Serrano 144, 28006 Madrid, Spain.}

\author{G. Chiappe}
\affiliation{Departamento de F\'{\i}sica Aplicada and Unidad
  Asociada del Consejo Superior de Investigaciones Cient\'{\i}ficas,
  Universidad de Alicante, San Vicente del Raspeig, Alicante 03690,
  Spain.}

\author{E. Louis}
\affiliation{Departamento de F\'{\i}sica Aplicada and Unidad
  Asociada del Consejo Superior de Investigaciones Cient\'{\i}ficas,
  Universidad de Alicante, San Vicente del Raspeig, Alicante 03690,
  Spain.}

\author{J.A. Miralles}
\affiliation{Departamento de F\'{\i}sica Aplicada and Unidad
  Asociada del Consejo Superior de Investigaciones Cient\'{\i}ficas,
  Universidad de Alicante, San Vicente del Raspeig, Alicante 03690,
  Spain.}

\author{J. A. Verg\'es}
\affiliation{Departamento de Teor\'{\i}a de la Materia Condensada,
  Instituto de Ciencia de Materiales de Madrid, Consejo Superior de
  Investigaciones Cient\'{\i}ficas, Cantoblanco, Madrid 28049, Spain.}

\author{J.L. Arag\'on\footnote{Electronic address:
    aragon@fata.unam.mx}}
\affiliation{Centro de F\'{\i}sica Aplicada y Tecnolog\'{\i}a
  Avanzada, Universidad Nacional Aut\'onoma de
  M\'exico, Apartado Postal 1-1010, Quer\'etaro 76000, M\'exico.}

\date{\today}

\begin{abstract}
  A subtle procedure to confine hydrodynamic modes on the free surface of a
  fluid is presented here. The experiment consists of a square vessel
  with an immersed square central well vibrating vertically so that
  the surface waves generated by the meniscus at the vessel boundary
  interfere with the bound states of the well. This is a classical
  analogy of a quantum well where some fundamental phenomena, such as
  bonding of states and interference between free waves and bound
  states, can be visualized and controlled. The above mentioned
  interference leads to a novel hydrodynamic transition from quasiperiodic
  to periodic patterns. Tight binding numerical calculations are performed
  here to show that our results could be transferred to design quantum
  confinements exhibiting electronic quasiperiodic surface states and their
  rational approximants for the first time.
\end{abstract}

\pacs{47.35.+i, 47.20.-k, 47.54.+r, 71.23.Ft}

\maketitle

The interest in experiments of classical analogs, which accurately
model the salient features of some quantum systems or other
fundamental undulatory phenomena, was firstly raised by the acoustic
experiments of Maynard \cite{Maynard}, where the analogy between
both the acoustic and Schr\"odinger equations is invoked. In a
general overview, a hydrodynamic analogy has also been used to
describe the flow of electrons in quantum semiconductor devices
\cite{Gardner}. Some experiments with liquid surface waves have been
reported later and they presented abstract concepts, such as Bloch
states, domain walls and band-gaps in periodic systems
\cite{Torres1,Torres2}, Bloch-like states in quasiperiodic systems
\cite{Torres3} or novel findings as the superlensing effect
\cite{Hu} in an visually clear manner. On the other hand, the
correspondence between the shallow water wave equation and the
acoustic wave equation has also been demonstrated
\cite{Chou,Torres2}. Such correspondences could be exploited to
investigate and address formally similar quantum effects as those
observed in quantum corrals \cite{Crommie}, where an optical analogy
has already been proposed \cite{Francs}, and in grain boundaries or
simple surface steps \cite{Hasegawa}. Our main goal is to build up
the hydrodynamic analogy of the interference between bound states in
a finite quantum well and free states. Then we realized an
experimental set up consisting of a square vessel with a single
square well drilled at its bottom; both squares are concentric and
well diagonals are parallel to the box sides. The immersed well was
expected to work as a weak potential binding surface waves upon
dependence on the liquid depth. When the vessel vibrated vertically
with such amplitudes that the Faraday instability was prevented,
such geometry produced three kinds of linear or weakly non-linear
patterns on the free surface of the liquid. The first pattern is a
sort of bound state restricted to the surface area occupied by the
immersed well that works as a weak potential binding standing plane
waves. The second pattern is produced by the meniscus at the vessel
walls \cite{Douady} and it can invade the region of the well
depending on the liquid depth and the vessel vibration frequency.
Finally, the last pattern arises from the interference between the
vessel meniscus waves and the bound states of the well. Patterns
observed depend on the liquid depth $h_1$, which plays the role of
an order parameter by controlling the amplitudes of the bound states
inside the immersed well.

Our main observation is summarized in Fig. \ref{fig:fig2}(a), which
clearly shows the binding of the surface wave produced by the
drilled well when the vibration amplitude is 60 $\mu$m. This pattern
will be detailed below but the physics of its origin can be
discussed as follows. The bound states arise from an inertial
hydrodynamic instability, balanced by the liquid surface tension
\cite{Landau}, that grows over the square well region
\cite{Torres3}. The bound state amplitudes increase on increasing $1
/ a^2$; where $a^2 = T/ \rho g$, $a$ is the capillary length, $T$ is
the liquid surface tension, $\rho$  is the liquid density and $g =
g_0 \pm \alpha \omega ^2$ is the effective gravity; $g_0$ is the
acceleration due to gravity, $\alpha$ is the vessel vibration
amplitude and $\omega$ is the vibration angular frequency of the
vessel.  On the contrary, the meniscus wave amplitude depends on the
variation of the meniscus volume during each vessel oscillation and,
it grows accordingly when $a^2$ increases \cite{Douady,Landau}. In
our experiment $\alpha$ is about 60 $\mu$m and the meniscus wave
amplitude reaches a maximum at a vessel vibration frequency of 64
Hz. On the other hand, the vessel vibration frequency represents the
wave state energy. The frequency and the wavelength are related
through the well known dispersion relation of the gravity-capillary
waves \cite{Torres1,Torres2,Torres3,Landau}. The present experiment
is essentially monochromatic as it occurs in the optical analogs
\cite{Francs} of quantum corrals. To show that Fig.
\ref{fig:fig1}(a) is not a Faraday pattern, we present a snapshot of
the system vibrating at about 70 $\mu$m in Fig. \ref{fig:fig1}(b),
when the Faraday instability is really triggered in the square well.
Figure \ref{fig:fig1}(b) shows a higher wavelength that matches with
the corresponding period doubling related to a Faraday wave pattern.

We chose a square vessel and the orientated square well
configuration to verify that the immersed well confined wave states.
Then we used a square methacrylate box with side $L$ of 8 cm where a
single square well with depth $d$ of 2 mm and side $l$ of 3.5 cm was
drilled at its bottom. The bottom of the vessel was covered with a
shallow ethanol layer of depth $h_1$. The liquid depth over the well
was then $h_2 = h_1 + d$.

As it was already mentioned, the vessel vibration amplitude was 60
$\mu$m, below the Faraday instability threshold set at about 70
$\mu$m. The vessel vibrated vertically at a single frequency lying
within the range from 35 to 60 Hz. An optimum frequency value was 50
Hz. Depending on the depth $h_1$, our experimental results can be
separated into three cases.

\textbf{Case I}. For $h_1$ lower than 1 mm, the experiment shows two
square lattices rotated 45$^o$ between them, namely the lattice of
the immersed square well is separated from the vessel square lattice
(Fig. \ref{fig:fig1}a). The external wave amplitudes are lower than
the wave amplitudes within the well and, furthermore, the external
wave reflection at the well step \cite{Lamb,Bartholomeusz,Miles} is
strong. For very shallow liquid layers, waves are only present
within the well at the vibration amplitudes of the experiment.

\textbf{Case II}. At a vessel vibration frequency of 50 Hz, when
$h_1$ is 1.2 mm, a quasicrystalline standing wave pattern appears
inside the immersed square region whereas the outer wave pattern is
a square network (Fig. \ref{fig:fig2}(a)). The immersed square well
works as a weak potential and binds standing plane waves with
eigenvectors $k_2 \mathbf{u} ^\prime _x$ and $k_2 \mathbf{u} ^\prime
_y$, parallel to the well sides. Nevertheless, it is transparent for
the standing waves of the square box which tunnel the immersed well
framework under the experimental conditions. Inside the square
region, the vessel eigenstates have eigenvectors $k_2 \mathbf{u}_x$
and $k_2 \mathbf{u}_y$, parallel to the outer box sides. Thus, the
standing state is finally described by $\psi_2 (k_2) + \psi ^\prime
_2 (k_2) = A_2 [\exp(\imath k_2 x) + \exp(\imath k_2 y)] + A^\prime
_2 [\exp(\imath k_2 x^\prime) + \exp( \imath k_2 y^\prime )]$ inside
the region of the square well. Equal phases along $x$ and $y$, and
$x^\prime$ and $y^\prime$, respectively, were assumed and $| A_2 | =
| A^\prime _2 |$ for the mentioned liquid depth and vibration
frequency. The interference of standing patterns $\psi_2$ with $\psi
^\prime _2$ increases the symmetry in the well from square
crystalline to octagonal quasicrystalline. Such pattern matches well
with the outer one described by $\psi_1 (k_1) = A_1 [\exp(\imath k_1
x) + \exp(\imath k_1 y)]$. According to the dispersion relation for
gravity-capillary waves described elsewhere
\cite{Torres1,Torres2,Torres3,Landau}, the difference between $k_1$
and $k_2$ is about 2\% and the refraction bending of about 1.1$^o$
at the boundary of the central window is negligible. Furthermore,
the external wave reflection at the well step
\cite{Lamb,Bartholomeusz,Miles} is also negligible with such
parameters. On the other hand, slender outgoing evanescent waves are
emitted at the boundary of the well and they play an important role
in the matching between patterns. It is important to remark that the
parameters of the experiment do not allow through the well known
dispersion relation \cite{Torres1,Torres2,Torres3,Landau} that the
wave inside the inner well were a subharmonic Faraday wave and the
outer wave were a harmonic meniscus wave. If this were the case,
then $k_1$ and $k_2$ should be very different.

\textbf{Case III}. When $h_1$ is increased, the potential of the
immersed well is seen increasingly weaker by the system, and
$A^\prime _2$ decreases accordingly. Under such conditions,
transitional patterns from a quasicrystalline form to a crystalline
one appear gradually on the hydrodynamic window.  Figure
\ref{fig:fig2}(b) shows a transitional pattern corresponding to a
liquid depth $h_1$ of 1.5 mm and an excitation frequency of 50 Hz.
According to crystallographic techniques of image processing
described elsewhere \cite{Torres3}, the Fourier transforms of the
experimental patterns are calculated (Figs. \ref{fig:fig2}(a) and
\ref{fig:fig2}(b)) and they are used to depict Fig. \ref{fig:fig3},
where the fast decay of $| A^\prime _2 / A_2 | ^2$ is shown on
increasing $h_1$. Figure \ref{fig:fig4} shows four frames of the
numerical simulation of the quasicrystal-crystal transition on the
hydrodynamic central square window. The complete movie is available
as Auxiliary Material.

The appearance of a quasiperiodic pattern inside the immersed square
region, in Case II, confirms that the square well is binding
standing plane waves. Such a pattern can only be produced by the
interference of two square patterns rotated by 45$^o$, namely the
pattern of bound standing waves of the square well is transparent to
the pattern of standing meniscus waves of the vessel if the adequate
conditions of the experimental parameters are fulfilled. On the
other hand, the square pattern of the vessel is only observed beyond
the region of the central square well. Moreover, there are
evanescent waves coming out of the boundary of the square well. The
observation of such waves also confirms the analogy with a quantum
well \cite{Messiah}.

The description of the observed standing wave patterns given in Case
II is the same as the used to describe the stationary states of a
particle moving in a potential given by a square well.  Using these
functions, we show a numerical simulation of the observed
quasicrystalline pattern in Fig. \ref{fig:fig3}(b). A trial and
error numerical method was used to fit at the boundaries and a
better matching is obtained if slender outgoing evanescent waves
emitted at the boundary of the well are considered.

To test the viability of a quantum scenario analogous to our
experimental results, we numerically studied the quantum confinement
of a double square well by using a tight-binding Hamiltonian in a $L
\times L$ cluster of the square lattice with a single atomic orbital
per lattice site \cite{Cuevas},
\begin{eqnarray*}
{\hat H} & = & \sum_{m,n} \epsilon_{m,n} | m ,n \rangle \langle m,n
| \\
 & & - \sum_{<mn;m'n'>} t_{m,n;m'n'} | m,n \rangle \langle m',n'|,
\end{eqnarray*}
where $|m,n \rangle$ represents an atomic orbital at site $(m,n)$,
and $\epsilon_{m,n}$ its energy. In order to simulate the inner
square we have explored several possibilities. Outside the inner
square, the energy of all orbitals are taken equal to zero. Besides,
the hopping energies between nearest-neighbor sites (the symbol $<>$
denotes that the sum is restricted to nearest-neighbors) were all
taken equal to 1. Instead, inside the inner square, we either
changed the orbital or the hopping energies. Some illustrative
results are shown in Fig. \ref{fig:fig5}. Figures \ref{fig:fig5}(a)
and (b) correspond to wavefunctions close to the band bottom,
\emph{i.e.}, long wavelengths and high linearity. In Fig.
\ref{fig:fig5}(a) the octagonal symmetry within the inner square is
clearly visible. Figure \ref{fig:fig5}(b) illustrates an effect that
is purely quantum, namely, the effect of the outer square is visible
even on a wavefunction localized in the inner square. Although one
cannot expect a one-to-one correspondence between the experiments
discussed here and this simple quantum simulation, the results
clearly suggest that similar effects could be observed in a suitable
quantum system. Figure \ref{fig:fig5}(c) shows a peculiar null
energy state which is not located at the band bottom. In this state
the hopping energy outside the square is equal to 1, whereas it is
0.7 inside the square and the general pattern conspicuously
resembles our experiment. The Fourier spectrum of the inner square
pattern has been performed and it corresponds to a rational
approximant of an octagonal quasicrystal generated by the following
vectors: $(2/ \sqrt{5}, 0)$, $(0, 2/ \sqrt{5})$, $2(2/ \sqrt{5}, 1/
\sqrt{5})$, $2(2/ \sqrt{5}, -1/ \sqrt{5})$, $2(1/ \sqrt{5}, 2/
\sqrt{5})$, $2(-1/ \sqrt{5}, 2/ \sqrt{5})$. Fourier spectra of Figs.
\ref{fig:fig5}(a) and (b) correspond to quasicrystalline states.
This suggests that the hopping energy could be a parameter to induce
quasicrystal-crystal transitions in confined quantum states.

In conclusion, we have described here a hydrodynamic experiment that
gives rise to the confinement of wave states on a free surface. It
constitutes a stirring macroscopic experimental scenario which
models some salient features of a quantum well and stimulates the
study and visualization of confined quasiperiodic quantum states for
the first time.

\begin{acknowledgments}
 Technical support from C. Zorrilla and S. Tehuacanero is gratefully acknowledged.
 This work has been partially supported by the Spanish MCYT  (BFM20010202 and
 MAT2002-04429), the Mexicans DGAPA-UNAM (IN-108502-3) and CONACyT (D40615-F),
 the Argentineans UBACYT (x210 and x447) and Fundaci\'on Antorchas, and the
 University of Alicante.
\end{acknowledgments}

\newpage

\begin{figure}[!t]
\caption{(a) Snapshot of the system below the Faraday instability
 onset, vibrating at 35 Hz with $h_1$ = 0.35 mm. Two clearly
 separated wave patterns are shown. The amplitude of the meniscus
 wave is much lower than that of the well-bound states. The
 reflexion coefficient at the well step is about 0.5.
 (b) Snapshot of the system at the onset of the Faraday instability
 in the square well. }
 \label{fig:fig1}
\end{figure}

\begin{figure}[!t]
\caption{(a) Snapshot of the system vibrating at 50 Hz for $h_1$ =
 1.2 mm. The well is transparent for the meniscus wave. The interference
 between such wave and the well-bound states gives rise to a perfect
 quasicrystalline pattern (Top). Fourier transform of the
 quasicrystalline wave pattern. Some secondary rings with a lower
 intensity appear. Such rings are due to weak nonlinearities
 corresponding to the parametrically driven experiment (Bottom). (b)
 Snapshot of the system vibrating at 50 Hz for $h_1$ = 1.5
 mm. The amplitude of the bound states is lower than that of the
 meniscus wave. A quasicrystal-crystal intermediate state appears on
 the well (Top). Fourier transform  of the transitional pattern. In
 the main octagonal ring there are four brighter spots (Bottom).}
 \label{fig:fig2}
\end{figure}

\begin{figure}[!t]
\caption{(a) Decay of  $| A^\prime _2 / A_2 | ^2$ when the depth of
 the liquid layer increases; the spots ($\blacklozenge$) on the
 figure are experimental points. (b) Simulation of the pattern in
 Fig. \ref{fig:fig2}(a) by means of a numerical trial and error
 fitting of the standing waves of the box and those waves
 bound by the square well, as well as the
 evanescent waves outgoing from the well boundary.}
 \label{fig:fig3}
\end{figure}

\begin{figure}[!t]
\caption{Four frames of the numerical simulation of the
 quasicrystal- crystal transition are shown as the liquid depth
 $h_1$ increases. The corresponding Fourier transforms are shown
 as insets. A complete movie is available as Auxiliary Material.}
 \label{fig:fig4}
\end{figure}

\begin{figure}[!t]
\caption{Quantum wavefunctions of the tight binding Hamiltonian
described in the text.}
 \label{fig:fig5}
\end{figure}

\end{document}